\documentclass[letter]{ieice}
\usepackage[dvips]{graphicx}
\field{}
\title{Local Modules in Imperative Languages}
\authorlist{
\authorentry{Keehang KWON}{m}{labelA}
\authorentry{Daeseong Kang}{n}{labelB}
}
\affiliate[labelA]{The authors are with Computer Eng., DongA University. email:khkwon@dau.ac.kr}
\affiliate[labelB]{The author is with Electronics Eng., DongA University.}
\received{2003}{1}{1}
\revised{2003}{1}{1}
\finalreceived{2003}{1}{1}

\makeatletter
\long\def\@makemyfntext#1{$^{\rm *}\ $ #1}

\long\def\@myfootnotetext#1{\insert\footins{\footnotesize
    \interlinepenalty\interfootnotelinepenalty 
    \splittopskip\footnotesep
    \splitmaxdepth \dp\strutbox \floatingpenalty \@MM
    \hsize\columnwidth \@parboxrestore
   \edef\@currentlabel{\csname p@footnote\endcsname\@thefnmark}\@makemyfntext
    {\rule{\z@}{\footnotesep}\ignorespaces
      #1\strut}}}

\def\myfootnotetext{\@ifnextchar
     [{\@xfootnotenext}{\xdef\@thefnmark{\thempfn}\@myfootnotetext}}
\makeatother

\newenvironment{numberedlist}
{\begin{list}{\makebox[20pt]{\hss(\arabic{itemno})\enspace}}
             {\usecounter{itemno}\labelwidth 20pt}}{\end{list}}

\newcounter{itemno}

\newcounter{itemno1}

\newcounter{itemno2}

\newcounter{exno}

\newcounter{defno}

\newenvironment{defn}{\refstepcounter{defno}\medskip \noindent {\bf
Definition \thedefno.\ }}{\medskip}

\newcommand{\sep}{\;\vert\;}

\newcommand{\oprove}{\vdash\kern-.6em\lower.7ex\hbox{$\scriptstyle O$}\,}

\newcommand{\Pscr}{{\cal P}}

\newcommand{\Mscr}{{\cal M}}

\newcommand{\Sscr}{{\cal S}}
\newcommand{\pderivation}{{\cal P}\kern -.1em\hbox{\rm -derivation}}
\newcommand{\pderivationl}{{\cal P}\kern -.1em\hbox{\em -derivation}}
\newcommand{\pderivable}{{\cal P}\kern -.1em\hbox{\rm -derivable}}
\newcommand{\pderivablel}{{\cal P}\kern -.1em\hbox{\em -derivable}}
\newcommand{\pderivations}{{\cal P}\kern -.1em\hbox{\rm -derivations}}
\newcommand{\pderivability}{{\cal P}\kern -.1em\hbox{\rm -derivability}}

\newcommand{\all}{\forall}

\newcommand{\ie}{{\em i.e.}}

\newsavebox{\lpartfig}
\newsavebox{\rpartfig}

\newenvironment{exmple}{
 \begingroup \begin{tabbing} \hspace{2em}\= \hspace{3em}\= \hspace{3em}\=
\hspace{3em}\= \hspace{3em}\= \hspace{3em}\= \kill}{
 \end{tabbing}\endgroup}

\newcommand{\lb}{\langle}
\newcommand{\rb}{\rangle}

\newcommand{\Ra}{\supset}  

     {\\* \hspace*{\fill} \end{trivlist}}

\newcommand{\muprolog}{{C$^{mod}$}}
\newcommand{\Ria}{\Rightarrow}

\begin{document}
\maketitle
\begin{summary}
We propose a  notion of {\it local} modules for imperative langauges.
 To be specific, we introduce a new {\it implication} statement of the form $D \Ra G$ 
 where $D$ is a module (\ie, a set of procedure declarations) and  $G$ is a statement.
 This statement tells the machine to add $D$ $temporarily$  to the program in the course of executing
 $G$. Thus, $D$ acts as a local module and will be $discarded$ after executing $G$.
It therefore provides  efficient   program management. In addition, we describe a new constructive
module language to improve code reuse. Finally, we describe a scheme which improves the heap
management in traditional languages.
 We illustrate our idea
via \muprolog, an extension of the core  C  with the new  statement.

\end{summary}
\begin{keywords}
local modules,  program management, memory management.
\end{keywords}

\section{Introduction}\label{sec:intro}

 Efficient program management in imperative  programming -- C, its extension \cite{KHP13} and Java -- is an 
important issue.  Yet this  has proven a difficult task, relying on  ad-hoc techniques such as
various cache/page replacement algorithms.

An analysis shows that this difficulty comes from the fact that the machine has no idea as to which modules are
to be used in the near future.
Therefore module management  can be made easier by making the programmer specify which modules are to be
used.  Toward this end, inspired by \cite{MNPS91,HM94,MN12}, we propose a new {\it implication} statement 
$D \Ra G$, 
where $D$ is a set of procedure declarations and   $G$ is a statement. This has the following execution semantics:
add $D$ temporarily to the current program in the course of executing $G$. 
In other words, the machine loads $D$ to the current program, executes $G$, and then unloads $D$ from
the current program.  This implication statement is closely related to the $let$-expression in
functional languages and the implication goals in logic languages \cite{MNPS91,MN12}.

Thus $D$ acts as $local$ procedures to $G$ in  that $D$ is hidden from the rest.
Our approach calls for a new runtime stack called {\it program stack},
 as it requires run-time loading and unloading. It follows that local procedures in our language is
(stack) dynamic scoped in the sense that the meaning of a procedure is always determined by
its most recent declaration.
On the other hand, most modern languages  allow local procedures within nested procedures.  
These approaches are based on static scoping in granting access and
 requires no run-time loading and unloading.
The main advantages of our approach is the following:

\begin{numberedlist}

\item It allows local procedures at the statement level, whereas other languages allow
local procedures only at the procedural level. Thus our langauge provides the
programmer more flexibility.

\item It  leads to efficient program/memory management due to loading and unloading.
     This is not negligible when local procedures are big.

\item It has a simple, natural syntax and semantics due to dynamic scoping of local procedures.
        In contrast, it is well-known that other  systems have awkward and
complicated semantics mainly due to   static  scoping.
Consequently, these systems are very difficult to read, write, implement  and reason about. 

\end{numberedlist}
On the negative side, it requires a little run-time overhead for loading and
unloading.

In the sequel, a module is nothing but a set of  procedures  with a name.
Our notion of local procedures extends to a notion of local (occurrences of )
modules in a straightforward way. That is, we propose a new {\it module implication} statement 
$/m \Ra G$, 
where $m$ is a module name and   $G$ is a statement. This has the following execution semantics:
add a (local occurrence of) module $m$ temporarily to the current program in the course of 
executing $G$. 
Note that our modules are  stack dynamic in the sense that they are loaded/unloaded in the program 
in a stack
fashion. This leads naturally to the dynamic scoping for procedure names.
In contrast, most  imperative languages have a  module language which
 is typically based on  the notion of static modules with no run-time loading and unloading,
 leading naturally to static scoping. It is well-known that static scoping causes the naming problem
among procedures across independent modules.

Our module system has some advantages over other popular module systems in imperative languages.

\begin{numberedlist}

\item It allows the programmer to load and unload other modules due to the module implication statement.
      This leads to the dynamic scoping for procedure names. 
      In contrast, this is traditionally impossible in other languages, leading to the static scoping
 and the naming problem.

\item  Dynamic scoping leads to  a simple, natural syntax and semantics, as there is no naming
problem.

\item As we shall see later, it allows mutually recursive modules thanks to dynamic scoping.

\end{numberedlist}

In addition, we add a novel module language to improve code reuse. 
Finally, we propose a variant of the implication statement which considerably 
simplifies the heap management.

This paper extends a C-like language with the new implication statement. We focus on the minimum 
core of C.
The remainder of this paper is structured as follows. We describe 
 \muprolog, an extension of  C  with a new 
 statement
 in Section 2. In Section \ref{sec:modules}, we
present an example of  \muprolog. In Section 4, we describe a constructive module language for 
enhancing code reuse. In Section 5, we propose a scheme that improves the heap management.
Section~\ref{sec:conc} concludes the paper.

\section{The Core Language}\label{sec:logic}

The language is   core C 
 with  procedure definitions. It is described
by $G$- and $D$-formulas given by the BNF syntax rules below:
\begin{exmple}
\>$G ::=$ \>   $true \sep A \sep x = E \sep  G;G \sep D \Ra G $ \\  
\>$D ::=$ \>  $ A = G\ \sep \all x\ D \sep D\land D$\\
\end{exmple}
\noindent
 In the above, $A$ 
 represents a head of a procedure declaration $p(x_1,\ldots,x_n)$ where $x_1,\ldots,x_n$ are parameters.
A $D$-formula is a  set
of procedure declarations.
In the transition system to be considered, a $G$-formula will function as a statement 
and a list of $D$-formulas  enhanced with the
machine state (a set of variable-value bindings) will constitute  a program.
Thus, a program is a pair of two disjoint components, \ie, $\lb D_1::\ldots::D_n::nil,\theta\rb$
where $D_1::\ldots::D_n::nil$ is a stack of  $D$-formulas and $\theta$ represents the machine state.
$\theta$ is initially  empty  and will be updated dynamically during execution
via the assignment statements. 

 We will  present an interpreter for our language via natural semantics \cite{Khan87}.
Note that  our interpreter  alternates between 
 the  execution phase 
and the backchaining phase.  
In  the  execution phase (denoted by $ex(\Pscr,G,\Pscr')$), it  
executes a statement $G$  with respect to
 $\Pscr$ and
produce a new program $\Pscr'$
by reducing $G$ 
to simpler forms. The rules
 (6)-(9) deal with this phase. 
If $G$ becomes a procedure call, the machine switches to the backchaining mode. This is encoded in the rule (5). 
In the backchaining mode (denoted by $bc(D,\Pscr,A,\Pscr')$), the interpreter tries 
to find a matching procedure  for a procedure call $A$ inside the module $D$
 by decomposing $D$ into a smaller unit (via rule (4)-(5)) and
 reducing $D$ to  its instance
 (via rule (2)) and then backchaining on the resulting 
definition (via rule (1)).
 To be specific, the rule (2) basically deals with argument passing: it eliminates the universal quantifier $x$ in $\all x D$
by picking a value $t$ for
$x$ so that the resulting instantiation,  $[t/x]D$, matches the procedure call $A$.
 The notation $S$\ seqand\ $R$ denotes the  sequential execution of two tasks. To be precise, it denotes
the following: execute $S$ and execute
$R$ sequentially. It is considered a success if both executions succeed.
Similarly, the notation $S$\ parand\ $R$ denotes the  parallel execution of two tasks. To be precise, it denotes
the following: execute $S$ and execute
$R$  in any order.  It is considered a success if both executions succeed.
The notation $S \leftarrow R$ denotes  reverse implication, \ie, $R \rightarrow S$.

\begin{defn}\label{def:semantics}
Let $G$ be a statement and let $\Pscr$ be the program.
Then the notion of   executing $\lb \Pscr,G \rb$ and producing a new
program $\Pscr'$-- $ex(\Pscr,G,\Pscr')$ --
 is defined as follows:

\begin{numberedlist}

\item    $bc((A = G_1),\Pscr,A,\Pscr_1)\ \leftarrow$  \\
 $ex(\Pscr,G_1,\Pscr_1)$. \% A matching procedure for $A$ is found.

\item    $bc(\all x D,\Pscr,A,\Pscr_1,)\ \leftarrow$  \\
  $bc([t/x]D,\Pscr, A,\Pscr_1)$. \% argument passing

\item    $bc( D_1\land D_2,\Pscr,A,\Pscr_1)\ \leftarrow$  \\
  $bc(D_1,\Pscr, A,\Pscr_1)$. \% look for  a matching procedure in $D_1$.

\item    $bc( D_1\land D_2,\Pscr,A,\Pscr_1)\ \leftarrow$  \\
  $bc(D_2,\Pscr, A,\Pscr_1)$. \% look for a matching procedure in $D_2$

\item    $ex(\Pscr,A,\Pscr_1)\ \leftarrow$    $(D_i \in \Pscr)$ parand $bc(D_i,\Pscr, A,\Pscr_1)$,
provided that $D_i$ is the first module in the stack,  which contains a declaration of $A$. 
\% $A$ is a procedure call

\item  $ex(\Pscr,true,\Pscr)$. \% True is always a success.



\item  $ex(\lb \Sscr,\theta\rb,x = E,\lb \Sscr,\theta \uplus \{ (x,E') \}\rb) \leftarrow$ $eval(\Pscr,E,E')$. \\
 \% evaluate $E$ to get $E'$.
 Here, 
$\uplus$ denotes a set union but $(x,V)$ in $\theta$ will be replaced by $(x,E')$.

\item  $ex(\Pscr,G_1; G_2,\Pscr_2)\ \leftarrow$ \\
  $ex(\Pscr,G_1,\Pscr_1)$  seqand  $ex(\Pscr_1,G_2,\Pscr_2)$.
\%  a sequential composition

\item  $ex(\lb \Sscr,\theta\rb,D \Ra G_1,\Pscr_1)\ \leftarrow$  \\
   $ex((\lb D::\Sscr,\theta \rb,G_1,\Pscr_1)$.  \% add $D$ to the top of the program stack $\Sscr$.

\item  $ex(\lb \Sscr,\theta\rb,/m \Ra G_1,\Pscr_1)\ \leftarrow$  \\
   $ex((\lb D::\Sscr,\theta \rb,G_1,\Pscr_1)$.  \% add $D$ to the top of the program stack $\Sscr$.

\end{numberedlist}
\end{defn}

\noindent
If $ex(\Pscr,G,\Pscr_1)$ has no derivation, then the interpreter returns  the failure.
The rule (9)  deals with the new feature.

\section{Examples}\label{sec:modules}

In our language, a module is simply a set of procedures associated with a name.
Below the keyword $module$ associates a name to a $D$-formula.
The following module Emp has a procedure Age which sets the variable named $age$,
 whose value represents the employee's age. 
Similarly, the module Bank is  defined 
with the procedures Deposit, Withdraw, Balance.

\begin{exmple}
module Emp.\\
Age(emp) =     \\
\>        switch (emp) \{ \\
 \>           case tom:  age = 31;   break; \\
  \>          case kim: age = 40;   break;\\
 \>           case sue: age = 22;    break;\\
 \>           default: age = 0;                     break;\\
        \}\\
\end{exmple}

\begin{exmple}
module Bank.\\
Deposit(name,amount) = $\ldots$    \\
Withdraw(name,amount) = $\ldots$     \\
Balance(name) = $\ldots$
\end{exmple}

\newpage

Now consider executing  the following main statement $G$ from an empty program.

\begin{exmple}
\% first task using module EmpAge   \\
\>      ( Emp $\Ria$   \\
 \>\>           (Age(tom); print(age); \\
  \>\>          Age(kim);  print(age);\\
 \>\>           Age(sue); print(age)))  \\
 \>      ;\\
\% second task using module Bank \\
\>   ( Bank  $\Ria$ \\
 \>\> deposit(tom,\$100)) 
\end{exmple}

\noindent Execution proceeds as follows: Initially the program is empty.
Then, the machine loads the module $Emp$ to the program,  printing the ages of 
three employees -- Tom, Kim and Sue --, and then unloads the module $Emp$.
Then, the machine loads the module $Bank$ to the program, deposits \$100 to Tom's account,
and then unloads the module $Bank$.
 Note that the module $Emp$ is available to the first task only, while $Bank$ 
to the second task only.

As the second example, let us consider two mutually recursive modules $Ev$ and $Od$.
The  module $Ev$ has a procedure Even(x) which returns true if x is even.
Similarly, the module $Od$ is  defined 
with the procedure Odd(x).

\begin{exmple}
module Ev.\\
Even(x) =     if x == 0, true else Od $\Ria$ Odd(x-1); 
\end{exmple}

\begin{exmple}
module Od.\\
Odd(x) =     if x == 1, true else Ev $\Ria$ Even(x-1); 
\end{exmple}


Now consider executing  even(9)  from the module Ev.
 Execution proceeds as follows: Initially the program is empty.
Then, the machine loads the module $Emp$ to the program,  printing the ages of 
three employees -- Tom, Kim and Sue --, and then unloads the module $Emp$.
Then, the machine loads the module $Bank$ to the program, deposits \$100 to Tom's account,
and then unloads the module $Bank$.
 Note that the module $Emp$ is available to the first task only, while $Bank$ 
to the second task only.

\section{A Constructive Module Langauge}\label{sec:intro}

Modern languages typically support code reuse via inheritance.
We propose a constructive approach to code reuse as an alternative to inheritance.
 To begin with,  our language provides a special macro function $/$
which binds a name to a set of method (and constant) declarations.
 This macro function serves to represent
programs in a concise way. For example, given two macro definitions
$/p\ =\ f(x) = x$
and   $/q\ =\ g(x) =  0$, the notation $/p \land /q$ represents $f(x) = x \land g(x) = 0$.
Here $\land$ means  the accumulation of two modules.

In addition to $\land$, our module language provides a 
renaming operation of the form $ren(b,a) D$ which replaces $b$ by $a$ in a module $D$
and $\all x D$ for universal generalization.
There are other useful operations such as $private\ f\ D$ (reuse $D$ with making $f$ private)
and $share D$ (reuse D via sharing, not copying) but we will not discuss them further here.

Now let us consider macro processing. Macro definitions are typically processed before the execution  but
in our setting, it is possible to process macros and execute regular programs in an
interleaved fashion.  We adopt this approach below.

We reconsider the language in Section 2.
 
\begin{exmple}
\>$G ::=$ \>   $true \sep A  \sep x = E \sep  G;G \sep D \Ra G \sep   /n:M \Ra  G $ \\  
\>$D ::=$ \>  $ A = G\ \sep /n \sep ren(a,b) D \sep \all x\ D \sep D\land D$\\
\>$M ::=$ \>  $ /n =  D \sep M \land M$\\
\end{exmple}
\noindent
 In the above, $n$ is a name and $A$ 
 represents a head of a procedure declaration $p(x_1,\ldots,x_n)$ where $x_1,\ldots,x_n$ are parameters.
A $D$-formula is a  set
of procedure declarations. An $M$-formula  is called  macro definitions and $\Mscr$ is a list of $M$-formulas.
In the transition system to be considered, a $G$-formula will function as a statement 
and a list of $D$-formulas,  a list of $M$-formulas and  the
machine state (a set of variable-value bindings) will constitute  a program.
Thus, a program is a pair of three disjoint components, \ie, $\lb D_1::\ldots::D_n::nil,\Mscr, \theta \rb$
where  $\theta$ represents the machine state.
$\theta$ is initially  empty  and will be updated dynamically during execution
via the assignment statements.

\begin{defn}\label{def:semantics}
Let $G$ be a statement and let $\Pscr$ be the program.
Then the notion of   executing $\lb \Pscr,G \rb$ and producing a new
program $\Pscr'$-- $ex(\Pscr,G,\Pscr')$ --
 is defined as follows:

\begin{numberedlist}

\item    $bc((A = G_1),\Pscr,A,\Pscr_1)\ \leftarrow$  \\
 $ex(\Pscr,G_1,\Pscr_1)$. \% A matching procedure for $A$ is found.

\item    $bc(\all x D,\Pscr,A,\Pscr_1,)\ \leftarrow$  \\
  $bc([t/x]D,\Pscr, A,\Pscr_1)$. \% argument passing

\item    $bc( D_1\land D_2,\Pscr,A,\Pscr_1)\ \leftarrow$  \\
  $bc(D_1,\Pscr, A,\Pscr_1)$. \% look for  a matching procedure in $D_1$.

\item    $bc( D_1\land D_2,\Pscr,A,\Pscr_1)\ \leftarrow$  \\
  $bc(D_2,\Pscr, A,\Pscr_1)$. \% look for a matching procedure in $D_2$

\item    $bc( ren(a,b) D,\Pscr,A,\Pscr_1)\ \leftarrow$  \\
  $bc([b/a]D,\Pscr, A,\Pscr_1)$. \% renaming operation

\item    $bc(/n,\Pscr,A,\Pscr_1)$ if  $bc(D,
\Pscr, A, \Pscr_1)$ and $(/n = D) \in \Mscr$.  \%  we assume it chooses the most recent macro definition.

\item    $ex(\Pscr,A,\Pscr_1)\ \leftarrow$    $(D_i \in \Pscr)$ parand $bc(D_i,\Pscr, A,\Pscr_1)$,
provided that $D_i$ is the first module in the stack,  which contains a declaration of $A$. 
\% $A$ is a procedure call

\item  $ex(\Pscr,true,\Pscr)$. \% True is always a success.



\item  $ex(\lb \Sscr,\Mscr,\theta\rb,x = E,\lb \Sscr,\Mscr,\theta \uplus \{ (x,E') \}\rb) \leftarrow$ $eval(\Pscr,E,E')$. \\
 \% evaluate $E$ to get $E'$.
 Here, 
$\uplus$ denotes a set union but $(x,V)$ in $\theta$ will be replaced by $(x,E')$.

\item  $ex(\Pscr,G_1; G_2,\Pscr_2)\ \leftarrow$ \\
  $ex(\Pscr,G_1,\Pscr_1)$  seqand  $ex(\Pscr_1,G_2,\Pscr_2)$.
\%  a sequential composition

\item  $ex(\lb \Sscr,\Mscr,\theta\rb,D \Ra G_1,\Pscr_1)\ \leftarrow$  \\
   $ex((\lb D::\Sscr,\Mscr,\theta \rb,G_1,\Pscr_1)$.  \% add $D$ to the top of the program stack $\Sscr$.

\item $ex(\lb \Sscr,\Mscr,\theta \rb, /n:M \Ra G_1,\Pscr_1)$ if  
$ex(\lb  /n::\Sscr, M::\Mscr, \theta\rb,G_1,\Pscr_1)$
\%  Add new macros to the front of $\Mscr$. Here $::$ is a list constructor.

\end{numberedlist}
\end{defn}

\noindent
If $ex(\Pscr,G,\Pscr_1)$ has no derivation, then the interpreter returns  the failure.

\section{Improving Heap Management}

Our earlier discussions in Section 2 are based on dynamic procedure binding.
More interestingly, our notion of 
implication statements can be applied equally well to static procedure/data binding.

To be specific, allocation and deallocation of  (heap) objects -- and accessing them through pointers --
occur frequently in traditional
imperative languages with static procedure/data binding.
 This includes malloc-free for memory management, new-dispose for objects,
new-delete for heap objects (arrays, records, etc). 
Unfortunately, allocation and deallocation constructs are unstructured. Using allocation and
deallocation carelessly leads to serious problems.

Towards an efficient yet robust memory management, we need to impose some
restrictions on the use of allocation and deallocation by providingng a high-level
statement. To be specific, we propose to use the following 
$pointer$-$implication$ statement of the form

\[ (p\ =\ {\rm new}\ obj) \supset G \]

where $p$ is a pointer and $obj$ is a program object or a data object (arrary, record, $etc$).
This is a variant of the implication statement in Section 1 with the following semantics:
It creates an object $obj$ with $p$ being a pointer to $obj$, executes the statement $G$ and then
deallocate $obj$ and $p$.
To avoid additional complications, we assume that pointers can only be initialized but not manipulated.

The following code is an example where arrayptr1 and int[100] are available in both the statements
$S_1$ and $S_2$,
while arrayptr2 and int[1000] are available only in  $S_2$.

\begin{exmple}
\% heap allocation   \\
\>      (arrayptr1 = new int[100] $\Ra$   \\
\>      (arrayptr2 = new int[1000] $\Ra$ $S_1$) \\
 \>\>          $S_2$) 
\end{exmple}

Our system requires considerable change to memory management: it
needs to maintain three different categories of memory: program/data stack, run-time stack and
the heap. Program/data stack is a new component and it -- instead of the heap -- 
is used to maintain program/data objects created via the $new$ construct. Thus it
replaces most of the works done by the heap.
Program/data stack has considerable advantages over the heap: it is more efficient and 
can simplify several complications caused by the heap including garbage collection,
heap fragmentation, and dangling pointers.

\section{Conclusion}\label{sec:conc}

In this paper, we have proposed a  simple  extension to imperative languages.
 This extension introduced an implication statement  $D \Ra G$ 
 where $D$ is a  module and $G$ is a statement. This statement makes $D$ local  to $G$. It 
therefore  maintains only the active modules in the current program context.

\section{Acknowledgements}

This work  was supported by Dong-A University Research Fund.

\bibliographystyle{ieicetr}



\end{document}